\documentclass[preprint,prd,onecolumn,preprintnumbers,superscriptaddress,showpacs,showkeys,nofootinbib]{revtex4-1}
\usepackage{graphicx,times}
\usepackage{amsmath,amsfonts,amssymb}
\usepackage{color}
\usepackage{psfrag}

\usepackage{hyperref}

\begin{document}
%\begin{CJK*}{GBK}{kai}¡ª
\title{Strong decays of $\bar{D}^{*}K^{*}$ molecules and the newly observed $X_{0,1}$ states }
\author{Yin Huang}
\affiliation{School of Physical Science and Technology, Southwest Jiaotong University, Chengdu 610031,China}

\author{Jun-Xu Lu}
\email{ljxwohool@buaa.edu.cn}
\affiliation{School of Physics, Beihang University, Beijing 100191, China}

\author{Jun-Jun Xie}\email{xiejujun@impcas.ac.cn}
\affiliation{Institute of Modern Physics, Chinese Academy of Sciences, Lanzhou 730000, China}
\affiliation{School of Nuclear Sciences and Technology, University of Chinese Academy of Sciences, Beijing 101408, China}
\affiliation{School of Physics and Microelectronics, Zhengzhou University, Zhengzhou, Henan 450001, China}

\author{Li-Sheng Geng}
\email{lisheng.geng@buaa.edu.cn}
\affiliation{School of Physics  \& Beijing Key Laboratory of Advanced Nuclear Materials and Physics, Beihang University, Beijing 100191, China}
\affiliation{Beijing Advanced Innovation Center for Big Date-based Precision Medicine, Beihang University, Beijing100191, China.}
\affiliation{School of Physics and Microelectronics, Zhengzhou University, Zhengzhou, Henan 450001, China}

\begin{abstract}
Lately, the LHCb Collaboration reported the discovery of two new states in the $B^+\rightarrow D^+D^- K^+$ decay, i.e., $X_0(2866)$ and $X_1(2904)$. In the present work, we study whether these states can be understood as $D^*\bar{K}^*$ molecules from the perspective of their two-body strong decays into $D^-K^+$ via triangle diagrams and three-body decays into $D^*\bar{K}\pi$. The coupling of the two states  to $D^*\bar{K}^*$ are determined from the Weinberg compositeness condition, while the
other relevant  couplings are well known. The obtained strong decay width for the $X_0(2866)$, in marginal agreement with
the experimental value within the uncertainty of the model, hints at a large $D^*\bar{K}^*$ component in its wave function. On the other hand, the strong decay width for the $X_1(2904)$,  much smaller than its experimental counterpart,  effectively rules out its assignment as  a $D^*\bar{K}^*$ molecule.
\end{abstract}

\date{\today}
\maketitle
%\CJKindent
%%%%%%%%%%%%%%%%%%%%%%%%%%%%%
\section{Introduction}
Ever since the experimental discovery of the $X(3872)$ and $D_{s0}^*(2317)$, many hadrons that cannot be simply classified into
conventional mesons of $q\bar{q}$ and baryons of $qqq$ have been discovered, with the latest addition being the $cc\bar{c}\bar{c}$ states discovered by the LHCb Collaboration~\cite{Aaij:2020fnh}. See, e.g., Refs.~\cite{Liu:2019zoy,Brambilla:2019esw,Guo:2017jvc,Liu:2019zoy,Chen:2016spr} for recent reviews. It should be noted that most of the so-called exotic hadrons mix with conventional hadrons or can be understood as hadron-hadron molecules or threshold effects such that they are not that  ``exotic''. Curiously, two of the truly exotic candidates, the $\theta^+(1540)$~\cite{Nakano:2003qx} and the $X(5568)$~\cite{D0:2016mwd} seem to fade away with time. In such a context, the latest LHCb
announcement of two structures observed in the $D^-K^+$ invariant mass of the $B^+ \rightarrow D^+D^-K^+$ decay points to the likely existence of
genuinely exotic mesonic states with a minimum quark content of $\bar{c}\bar{s}ud$~\cite{lhcb-x2900}. Their masses and widths are, in units of MeV, respectively
\begin{equation}
X_0(2866):\quad M=2866\pm7 \quad\mbox{and}\quad \Gamma=57.2\pm12.9,
\end{equation}
\begin{equation}
X_1(2900):\quad M=2904\pm5 \quad\mbox{and}\quad\Gamma=110.3\pm11.5.
\end{equation}
The spin-parities of these two states are determined to be $0^+$ and $1^-$.

It is interesting to note that these two states are just below ($X_0$) and close to ($X_1$) the $D^*\bar{K}^*$ threshold. Although
the existence of compact tetraquark states in this energy region has been predicted, in either quark models~\cite{Cheng:2020nho,Tan:2020ldi,Liu:2016ogz,Lu:2016zhe}, or QCD sum rules~\cite{Chen:2017rhl,Tang:2016pcf}~\footnote{It is interesting to note that a state of the art lattice QCD study found no compact tetraquark state of $\bar{c}\bar{s}ud$ with $I=0$ and spin-parity $0^+$ and
$1^+$~\cite{Hudspith:2020tdf}.}. In the present work, we examine the possibility whether they can be understood as $D^*\bar{K}^*$ molecules. For such a purpose, we first assume that they are bound states of $D^*\bar{K}^*$, and then employ the weinberg compositeness rule to determine their couplings to $D^*\bar{K}^*$. The two body strong decays then follow from the exchange of a pseudoscalar meson between the $D^*\bar{K}^*$ pair, which then transforms into $D^-K^+$. Such a process is depicted in Fig.~1. In addition, the $D^*\bar{K}^*$ molecules can also
 decay into a three-body finale state $D^*\bar{K}\pi$, as shown in Fig.~2.\footnote{As the $D^*$ is very narrow, we treat it as a stable particle.} If within the uncertainties of the model, the so-obtained strong decay widths are consistent with data, then it is possible to assign the state under study as a molecular state, otherwise, the possibility is excluded. Such an approach has been widely applied to study newly observed (exotic) hadrons, see, e.g., Refs.~\cite{Huang:2019qmw,Faessler:2007us,Dong:2009yp,Dong:2009uf,Dong:2017gaw,Xiao:2019mst,Huang:2018wgr,Huang:2018bed} for a partial list.

It is interesting to note that the  $DDK$ bound state of isospin $1/2$ and spin-parity $0^-$ with a mass around 4140 MeV~\cite{SanchezSanchez:2017xtl,MartinezTorres:2018zbl,Wu:2019vsy} is different from those observed by the LHCb Collaboration in the $D^-K^+$ spectrum. Though the former is built from the $DDK$ interaction, it decays into $DD_s\pi$~\cite{Huang:2019qmw} instead of $D^-K^+$ because of parity conservation. It would be interesting  if in the future the LHCb collaboration can search for the existence of such a state.

This work is organized as follows. In Section II, we explain the theoretical formalism. Results and discussions
are provided in Section III, followed by a short summary in Section IV.

\section{Theoretical framework}
\begin{figure}[htbp]
\begin{center}
\includegraphics[scale=0.65]{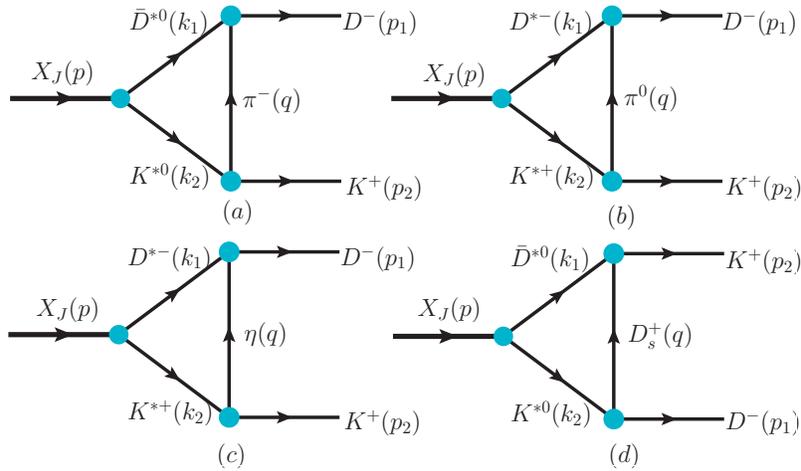}
\caption{Diagrams representing the decay of the $X_{J=0,1}$ states to $D^{-}K^{+}$.}\label{mku}
\end{center}
\end{figure}

\begin{figure}[htbp]
\begin{center}
\includegraphics[scale=0.65]{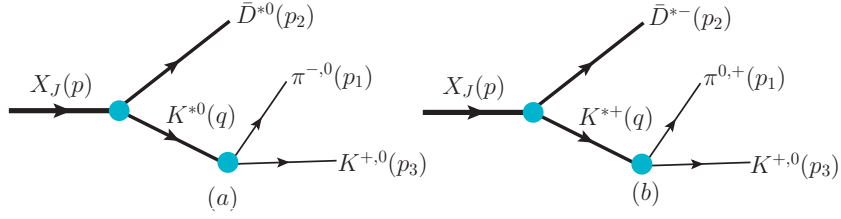}
\caption{Diagrams representing the decay of the $X_J$ state to $\bar{D}^{*}K\pi$.}\label{mkuds}
\end{center}
\end{figure}

In the following, we explain how the strong decays into $DK$, Fig.~\ref{mku}, and $D^*K\pi$, Fig.~\ref{mkuds}, are computed. We take advantage of the fact that the $D^*$ is very narrow (with a width of less than 100 keV) and therefore can be treated as a stable particle for our purpose.

We shall construct the amplitudes using the isospin formalism,  where the $\bar{D}K^*$ isospin doublet reads
\begin{align}
|K^{*}\bar{D}^{*},I=0\rangle&=\frac{1}{\sqrt{2}}(K^{*+}D^{*-}+K^{*0}\bar{D}^{*0}),\\
|K^{*}\bar{D}^{*},I=1\rangle&=\frac{1}{\sqrt{2}}(K^{*+}D^{*-}-K^{*0}\bar{D}^{*0})
\end{align}

Considering quantum numbers and phase space, the two body strong decay modes of $X_J$ are $X_J\to{}D^{-}K^{+}$ and
$X_J\to{}\bar{D}^{0}K^{0}$.  In this work, we only explicitly compute the partial
decay width of $X_J\to{}D^{-}K^{+}$, and that of $X_J\to{}\bar{D}^{0}K^{0}$ can be obtained by isospin symmetry
$\Gamma_{X_J\to{}D^{-}K^{+}}=\Gamma_{X_J\to{}\bar{D}^{0}K^{0}}$.  The sum of the two parts is the total
decay width of the $X_J\to\bar{D}K$.

In order to calculate the Feynman diagrams shown in Fig.~\ref{mku}, we need to determine  the
relevant vertices. For the vertex of $X_J\bar{D}^{*}K^{*}$, since the $X_J$ is
considered as a bound state of $\bar{D}^*K^{*}$, this coupling can be determined by the Weinberg
compositeness condition. In the present work, we adopt the method developed in Refs.~\cite{Huang:2019qmw,Faessler:2007us,Dong:2009yp,Dong:2009uf,Dong:2017gaw,Xiao:2019mst,Huang:2018wgr,Huang:2018bed}.
In this framework, the relevant Lagrangians for the  $X_0(2866)$
can be written as~\cite{Faessler:2007us}
\begin{figure}[htbp]
\begin{center}
\includegraphics[scale=0.8]{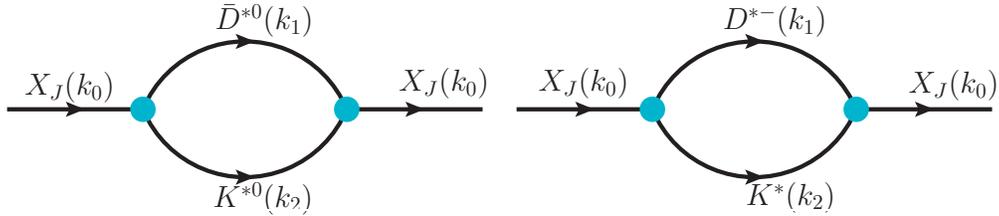}
\caption{Mass operators of the $X_J$ states.}\label{mku1}
\end{center}
\end{figure}

\begin{align}
{\cal{L}}_{X_0}(x)=g_{X_0\bar{D}^{*}K^{*}}X_J(x)\int{}dy\Phi(y^2)\bar{D}^{*\mu}(x+\omega_{K^{*}}y)K^{*}_{\mu}(x-\omega_{\bar{D}^{*}}y)+H.c.,\label{eq1}
\end{align}
while for the $X_1(2904)$ the Lagrangian has the form~\cite{Chen:2013bha}
\begin{align}
{\cal{L}}_{X_1}(x)&=g_{X_1\bar{D}^{*}K^{*}}X_J^{\alpha}(x)\int{}dy\Phi(y^2)\bar{D}^{*}_{\mu}(x+\omega_{K^{*}}y) \overleftrightarrow{\partial}_{\alpha}K^{*\mu}(x-\omega_{\bar{D}^{*}}y).\label{eq2}
\end{align}
where $\omega_{i}=m_{i}/(m_i+m_j)$  is a kinematical parameter with $m_i$ and $m_j$ being the masses of the involved mesons.
In the Lagrangians, an effective correlation function $\Phi(y^2)$ is introduced to describe the distribution of the two constituents,
$\bar{D}^{*}$ and $K^{*}$, in the hadronic molecular $X_J$ states.  The introduced correlation function also serves the purpose of making the
Feynman diagrams ultraviolate finite. Here we choose the Fourier transformation of the
correlation function to have a Gaussian form,
\begin{align}
\Phi(-p_E^2)\doteq\exp(-p_E^2/\alpha^2),
\end{align}
where $\beta$ being the size parameter which characterizes the distribution
of the constituents inside the molecule.
The value of $\alpha$  has to be determined by fitting to data.  It is found that
the experimental total decay widths of some states that
can be considered as molecules (see, e.g., Refs.~\cite{Huang:2019qmw,Faessler:2007us,Dong:2009yp,Dong:2009uf,Dong:2017gaw,Xiao:2019mst,Huang:2018wgr,Huang:2018bed} and references therein) can be well explained with $\alpha\approx1.0$ GeV.   Therefore we take $\alpha=1.0\pm0.1$
GeV in this work to study whether the $X_J$ states can be interpreted as molecules composed of $\bar{D}^{*}K^{*}$.

The coupling constant $g_{X_J\bar{D}^*K^{*}}$ is determined by the compositeness condition~\cite{Huang:2019qmw,Faessler:2007us,Dong:2009yp,Dong:2009uf,Dong:2017gaw,Xiao:2019mst,Huang:2018wgr,Huang:2018bed}.
It implies that the renormalization constant of the hadron wave function is set to zero, i.e.,
\begin{align}
Z_{X_J}=1-\frac{d\Sigma^{~~T}_{0,1}}{dk\!\!\!/_0}|_{k\!\!\!/_0=m_{X_J}}=0\label{eqn3},
\end{align}
The $\Sigma^T_{1}$ is the transverse part of the self-energy operator $\Sigma^{\mu\nu}_{1}$, related to $\Sigma^{\mu\nu}_{1}$ via
\begin{align}
\Sigma^{\mu\nu}_{1}(k_0)=(g_{\mu\nu}-\frac{k_0^{\mu}k_0^{\nu}}{k_0^2})\Sigma^{T}_{1}+\cdots.\label{eqn4}
\end{align}
The concrete forms of the mass operator of the $X_J$  corresponding to Fig.~\ref{mku1} are
\begin{align}
\Sigma_0(k_0)&=\sum_{Y=\bar{D}^{0}K^{0},D^{*-}K^{*+}}({\cal{C}}^T_{Y})^2g^2_{X_0\bar{D}^{*}K^{*}}\int\frac{d^4k_1}{(2\pi)^4}\Phi^2[(k_1-k_0\omega_{\bar{D}^{*}})^2]\nonumber\\
                                &\times\frac{-g^{\mu\nu}+k_1^{\mu}k_1^{\nu}/m^2_{\bar{D}^{*}}}{k_1^2-m^2_{\bar{D}^{*}}}\frac{-g^{\mu\nu}+(k_0-k_1)^{\mu}(k_0-k_1)^{\nu}/m^2_{K^{*}}}{(k_0-k_1)^2-m^2_{K^{*}}},
\end{align}
\begin{align}
\Sigma^{\alpha\beta}_{1}&(k_0)=\sum_{Y=\bar{D}^{0}K^{0},D^{*-}K^{*+}}({\cal{C}}^T_{Y})^2g^2_{X_1\bar{D}^{*}K^{*}}\int\frac{d^4k_1}{(2\pi)^4}\Phi^2[(k_1-k_0\omega_{\bar{D}^{*}})^2]\nonumber\\
                                          &\times[k_1^{\alpha}k_1^{\beta}-k_1^{\alpha}(k_0-k_1)^{\beta}-k_1^{\beta}(k_0-k_1)^{\alpha}+(k_0-k_1)^{\alpha}(k_0-k_1)^{\beta}]\nonumber\\
                                           &\times\frac{-g^{\mu\nu}+k_1^{\mu}k_1^{\nu}/m^2_{\bar{D}^{*}}}{k_1^2-m^2_{\bar{D}^{*}}}\frac{-g^{\mu\nu}+(k_0-k_1)^{\mu}(k_0-k_1)^{\nu}/m^2_{K^{*}}}{(k_0-k_1)^2-m^2_{K^{*}}},\nonumber\\
\end{align}
where $z=2+\alpha+\beta$, $\Delta=-4\omega_{\bar{D}^{*}}k_0-2\beta{}k_0$, and $k_0^2=m^2_{X}$ with $k_0$, $m_{X}$ denoting the four-momenta and mass of the $X_J$, respectively.  Here, we set $m_{X_J}=m_{\bar{D}^{*}}+m_{K^{*}}-E_b$ with $E_b$ the binding energy of $X_J$, $k_1$, and $m_{\bar{D}^{*}}$ are  the four-momenta and mass of the $\bar{D}^{*}$, and $m_{K^{*}}$ is the  mass of $K^{*}$, respectively.  $I$ is isospin and isospin symmetry implies that
\begin{align*}
\begin{split}
{\cal{C}}^{I=0}_Y= \left \{
\begin{array}{ll}
    1/\sqrt{2},                    & Y=\bar{D}^{0}K^{0}\\
    1/\sqrt{2},                    & Y=D^{*-}K^{*+}
\end{array}
\right. ,
\end{split}
\end{align*}
and
\begin{align*}
\begin{split}
{\cal{C}}^{I=1}_Y= \left \{
\begin{array}{ll}
    -1/\sqrt{2},                    & Y=\bar{D}^{0}K^{0}\\
    1/\sqrt{2},                     & Y=D^{*-}K^{*+}
\end{array}
\right. .
\end{split}
\end{align*}

To evaluate the diagrams of Fig.~\ref{mku} and Fig.~\ref{mkuds}, in addition to the Lagrangians in Eqs.(~\ref{eq1},\ref{eq2}), the following effective
Lagrangians, responsible for the interaction between a vector meson and a pseudoscalar meson,  are needed
as well~\cite{Hofmann:2005sw}
\begin{align}
{\cal{L}}_{PPV}=\frac{i}{4}g_{h}\langle[\partial^{\mu}P,P]V_{\mu}\rangle\label{eq9},
\end{align}
where $P$ and  $V_{\mu}$ represents the vector fields of the 16-plet of the $\rho$ and the $SU(4)$ pseudoscalar meson matrix, respectively.
The $\langle...\rangle$ denotes  trace in the $SU(4)$ flavor space.  The meson matrices are~\cite{Hofmann:2005sw}
\begin{equation}
P=\sqrt{2}
\left(
  \begin{array}{cccc}
    \frac{\pi^{0}}{\sqrt{2}}+\frac{\eta}{\sqrt{6}}+\frac{\eta^{'}}{\sqrt{3}}   &\pi^{+}    &K^{+}      & \bar{D}^0    \\
    \pi^{-}  &  -\frac{\pi^{0}}{\sqrt{2}}+\frac{\eta}{\sqrt{6}}+\frac{\eta^{'}}{\sqrt{3}}  &K^{0}      &-D^{-}  \\
     K^{-}   &    \bar{K}^{0}  & -\sqrt{\frac{2}{3}}\eta+\frac{\eta^{'}}{\sqrt{3}}                     & D^{-}_s\\
     D^0     &    -D^{+}       &D_s^{+}                                                                &\eta_c   \\
  \end{array}
\right)\label{eqwq9}
\end{equation}
and
\begin{equation}
V_{\mu}=
\left(
  \begin{array}{cccc}
    \frac{1}{\sqrt{2}}(\rho^{0}+\omega) & \rho^{+}                             &  K^{*+}     & \bar{D}^{*0} \\
    \rho^{-}                            & \frac{1}{\sqrt{2}}(-\rho^{0}+\omega) &  K^{*0}     & D^{*-}       \\
     K^{*-}                             & \bar{K}^{*0}                         &  \phi       & D^{*-}_s     \\
     D^{*0}                             & D^{*+}                               &  D^{*+}_{s} & J/\psi       \\
  \end{array}
\right)_{\mu}.
\end{equation}
Then we obtain
\begin{align}
{\cal{L}}_{\pi{}DD^{*}}&=\frac{ig_h}{2\sqrt{2}}(\pi^{0}\partial^{\mu}D^{+}-D^{+}\partial^{\mu}\pi^0)\bar{D}^{*-}_{\mu}-\frac{ig_h}{2}(\pi^{-}\partial^{\mu}D^{+}-D^{+}\partial^{\mu}\pi^{-})\bar{D}^{*0}_{\mu}\nonumber\\
                       &+\frac{ig_h}{2}(\pi^{+}\partial^{\mu}D^{0}-D^{0}\partial^{\mu}\pi^+)\bar{D}^{*-}_{\mu}+\frac{ig_h}{2\sqrt{2}}(\pi^{0}\partial^{\mu}D^{0}-D^{0}\partial^{\mu}\pi^0)\bar{D}^{*0}_{\mu}\nonumber\\
                       &-\frac{ig_h}{2\sqrt{2}}(\pi^{0}\partial^{\mu}\bar{D}^{-}-\bar{D}^{-}\partial^{\mu}\pi^0)D^{*+}_{\mu}+\frac{ig_h}{2}(\pi^{+}\partial^{\mu}\bar{D}^{-}-\bar{D}^{-}\partial^{\mu}\pi^+)D^{*0}_{\mu}\nonumber\\
                       &-\frac{ig_h}{2}(\pi^{-}\partial^{\mu}\bar{D}^{0}-\bar{D}^{0}\partial^{\mu}\pi^-)D^{*+}_{\mu}-\frac{ig_h}{2\sqrt{2}}(\pi^{0}\partial^{\mu}\bar{D}^{0}-\bar{D}^{0}\partial^{\mu}\pi^0)D^{*0}_{\mu},
\end{align}
\begin{align}
{\cal{L}}_{\eta{}DD^{*}}&=-\frac{ig_h}{2\sqrt{6}}(\eta\partial^{\mu}D^{+}-D^{+}\partial^{\mu}\eta)\bar{D}^{*-}_{\mu}+\frac{ig_h}{2\sqrt{6}}(\eta\partial^{\mu}D^{0}-D^{0}\partial^{\mu}\eta)\bar{D}^{*0}_{\mu}\nonumber\\
                       &+\frac{ig_h}{2\sqrt{6}}(\eta\partial^{\mu}\bar{D}^{-}-\bar{D}^{-}\partial^{\mu}\eta)D^{*+}_{\mu}-\frac{ig_h}{2\sqrt{6}}(\eta\partial^{\mu}\bar{D}^{0}-\bar{D}^{0}\partial^{\mu}\eta)D^{*0}_{\mu},
\end{align}
\begin{align}
{\cal{L}}_{\pi{}KK^{*}}&=-\frac{ig_h}{2\sqrt{2}}(\pi^{0}\partial^{\mu}K^{+}-K^{+}\partial^{\mu}\pi^0)\bar{K}^{*-}_{\mu}-\frac{ig_h}{2}(\pi^{-}\partial^{\mu}K^{+}-K^{+}\partial^{\mu}\pi^{-})\bar{K}^{*0}_{\mu}\nonumber\\
                       &-\frac{ig_h}{2}(\pi^{+}\partial^{\mu}K^{0}-K^{0}\partial^{\mu}\pi^+)\bar{K}^{*-}_{\mu}+\frac{ig_h}{2\sqrt{2}}(\pi^{0}\partial^{\mu}K^{0}-K^{0}\partial^{\mu}\pi^0)\bar{K}^{*0}_{\mu}\nonumber\\
                       &+\frac{ig_h}{2\sqrt{2}}(\pi^{0}\partial^{\mu}\bar{K}^{-}-\bar{K}^{-}\partial^{\mu}\pi^0)K^{*+}_{\mu}+\frac{ig_h}{2}(\pi^{+}\partial^{\mu}\bar{K}^{-}-\bar{K}^{-}\partial^{\mu}\pi^+)K^{*0}_{\mu}\nonumber\\
                       &+\frac{ig_h}{2}(\pi^{-}\partial^{\mu}\bar{K}^{0}-\bar{K}^{0}\partial^{\mu}\pi^-)K^{*+}_{\mu}-\frac{ig_h}{2\sqrt{2}}(\pi^{0}\partial^{\mu}\bar{K}^{0}-\bar{K}^{0}\partial^{\mu}\pi^0)K^{*0}_{\mu},\label{eq15}
\end{align}
\begin{align}
{\cal{L}}_{\eta{}KK^{*}}&=-i\frac{\sqrt{6}g_h}{4}(\eta\partial^{\mu}K^{+}-K^{+}\partial^{\mu}\eta)\bar{K}^{*-}_{\mu}-i\frac{\sqrt{6}g_h}{4}(\eta\partial^{\mu}K^{0}-K^{0}\partial^{\mu}\eta)\bar{K}^{*0}_{\mu}\nonumber\\
                       &+i\frac{\sqrt{6}g_h}{4}(\eta\partial^{\mu}\bar{K}^{-}-\bar{K}^{-}\partial^{\mu}\eta)K^{*+}_{\mu}+i\frac{\sqrt{6}g_h}{4}(\eta\partial^{\mu}\bar{K}^{0}-\bar{K}^{0}\partial^{\mu}\eta)K^{*0}_{\mu},
\end{align}
\begin{align}
{\cal{L}}_{D^{*}D_sK}&=-\frac{ig_h}{2}(K^0\partial^{\mu}D_s^{-}-D_s^{-}\partial^{\mu}K^0)D^{*+}_{\mu}-\frac{ig_h}{2}(K^+\partial^{\mu}D_s^{-}-D_s^{-}\partial^{\mu}K^+)D^{*0}_{\mu}\nonumber\\
                       &+\frac{ig_h}{2}(\bar{K}^0\partial^{\mu}D_s^{+}-D_s^{+}\partial^{\mu}\bar{K}^0)\bar{D}^{*-}_{\mu}+\frac{ig_h}{2}(K^-\partial^{\mu}D_s^{+}-D_s^{+}\partial^{\mu}K^-)\bar{D}^{*0}_{\mu},
\end{align}
\begin{align}
{\cal{L}}_{DD_sK^*}&=-\frac{ig_h}{2}(D^+\partial^{\mu}D_s^{-}-D_s^{-}\partial^{\mu}D^+)K^{*0}_{\mu}-\frac{ig_h}{2}(\bar{D}^0\partial^{\mu}D_s^{+}-D_s^{+}\partial^{\mu}\bar{D}^0)K^{*-}_{\mu}\nonumber\\
                       &+\frac{ig_h}{2}(D^0\partial^{\mu}D_s^{-}-D_s^{-}\partial^{\mu}D^0)K^{*+}_{\mu}+\frac{ig_h}{2}(D^-\partial^{\mu}D_s^{+}-D_s^{+}\partial^{\mu}D^-)\bar{K}^{*0}_{\mu}.
\end{align}
The coupling $g_h$ is fixed from the strong decay width of $K^{*}\to{}K\pi$.  With the help of Eq.~(\ref{eq15}), the two-body decay width
 $\Gamma(K^{*+}\to{}K^{0}\pi^{+})$ is related to $g_h$ as
 \begin{align}
 \Gamma(K^{*+}\to{}K^{0}\pi^{+})=\frac{g_h^2}{24\pi{}m^2_{K^{*+}}}{\cal{P}}^3_{\pi{}K^{*}}=\frac{2}{3}\Gamma_{K^{*+}},
 \end{align}
where ${\cal{P}}_{\pi{}K^{*}}$ is the three-momentum of the $\pi$ in the rest frame of the  $K^{*}$.
Using the experimental strong decay width($\Gamma_{K^{*+}}=50.3\pm0.8$ MeV) and the masses of the particles  listed in Table~\ref{table1}~\cite{PAZylaetal2020}, we obtain $g_h=9.11$.
\begin{table}[h!]
\centering
\caption{Masses of the particles needed in the present work (in units of MeV).}\label{table1}
\begin{tabular}{cccccccccc}
\hline\hline
~~~   $D^{*0}$      ~~~ &$D^{*\pm}$      ~~~&$\eta$           ~~~ &$D_s^{\pm}$ ~~~  &$D^{0}$ ~~~     & $D^{\pm}$          \\
~~~   $2006.85$     ~~~ &$2010.26$      ~~~  &$547.86$       ~~~ &$1968.34$        ~~~  &$1864.83$    ~~~  & $1869.65   $        \\ \hline
~~~   $K^{0}$       ~~~ &$K^{*0}$       ~~~  &$K^{*\pm}$     ~~~ &$K^{\pm}$       ~~~  &$\pi^{\pm}$ ~~~  & $\pi^{0}$            \\
~~~   $497.611$     ~~~ &$898.36$      ~~~  &$891.66$       ~~~ &$493.68$        ~~~  &$139.57$    ~~~  & $134.98$              \\  \hline  \hline
\end{tabular}
\end{table}

\subsection{Two-body decay width}
With the above formalism, the decay amplitudes of the triangle diagrams of Fig.~\ref{mku}, evaluated in the final state center of mass frame, are
\begin{align}
{\cal{M}}^{X_J}_{a}&=i^3\frac{g_h^2g_{X_J\bar{D}^{*}K^{*}}}{4}{\cal{C}}_Y^{I}\int\frac{d^4q}{(2\pi)^4}\Phi[(k_1\omega_{K^{*0}}-k_2\omega_{\bar{D}^{*0}})^2]\nonumber\\
                          &\times(p_1^{\mu}+q^{\mu})(q^{\eta}-p^{\eta}_{2})\{1,i(k_2^{\alpha}-k_1^{\alpha})\epsilon^{X}_{\alpha}\}\nonumber\\
                          &\times\frac{-g^{\mu\nu}+k^{\mu}_{1}k^{\nu}_{1}/m^2_{\bar{D}^{*0}}}{k_1^2-m^2_{\bar{D}^{*0}}}\frac{-g^{\nu\eta}+k^{\nu}_{2}k^{\eta}_{2}/m^2_{K^{*0}}}{k_2^2-m^2_{K^{*0}}}\frac{1}{q^2-m^2_{\pi^{-}}},\label{eq22}\\
{\cal{M}}^{X_J}_{b}&=-i^3\frac{g_h^2g_{X_J\bar{D}^{*}K^{*}}}{8}{\cal{C}}_Y^{I}\int\frac{d^4q}{(2\pi)^4}\Phi[(k_1\omega_{K^{*+}}-k_2\omega_{D^{*-}})^2]\nonumber\\
                          &\times(p_1^{\mu}+q^{\mu})(q^{\eta}-p^{\eta}_{2})\{1,i(k_2^{\alpha}-k_1^{\alpha})\epsilon^{X}_{\alpha}\}\nonumber\\
                          &\times\frac{-g^{\mu\nu}+k^{\mu}_{1}k^{\nu}_{1}/m^2_{D^{*-}}}{k_1^2-m^2_{D^{*-}}}\frac{-g^{\nu\eta}+k^{\nu}_{2}k^{\eta}_{2}/m^2_{K^{*+}}}{k_2^2-m^2_{K^{*+}}}\frac{1}{q^2-m^2_{\pi^{0}}},\\
{\cal{M}}^{X_J}_{c}&=i^3\frac{g_h^2g_{X_J\bar{D}^{*}K^{*}}}{8}{\cal{C}}_Y^{I}\int\frac{d^4q}{(2\pi)^4}\Phi[(k_1\omega_{K^{*+}}-k_2\omega_{D^{*-}})^2]\nonumber\\
                          &\times(p_1^{\mu}+q^{\mu})(q^{\eta}-p^{\eta}_{2})\{1,i(k_2^{\alpha}-k_1^{\alpha})\epsilon^{X}_{\alpha}\}\nonumber\\
                          &\times\frac{-g^{\mu\nu}+k^{\mu}_{1}k^{\nu}_{1}/m^2_{D^{*-}}}{k_1^2-m^2_{D^{*-}}}\frac{-g^{\nu\eta}+k^{\nu}_{2}k^{\eta}_{2}/m^2_{K^{*+}}}{k_2^2-m^2_{K^{*+}}}\frac{1}{q^2-m^2_{\eta}},\\
{\cal{M}}^{X_J}_{d}&=i^3\frac{g_h^2g_{X_J\bar{D}^{*}K^{*}}}{4}{\cal{C}}_Y^{I}\int\frac{d^4q}{(2\pi)^4}\Phi[(k_1\omega_{K^{*0}}-k_2\omega_{\bar{D}^{*0}})^2]\nonumber\\
                          &\times(p_2^{\eta}+q^{\eta})(q^{\mu}-p^{\mu}_{1})\{1,i(k_2^{\alpha}-k_1^{\alpha})\epsilon^{X}_{\alpha}\}\nonumber\\
                          &\times\frac{-g^{\nu\eta}+k^{\nu}_{1}k^{\eta}_{1}/m^2_{\bar{D}^{*0}}}{k_1^2-m^2_{\bar{D}^{*0}}}\frac{-g^{\mu\nu}+k^{\mu}_{2}k^{\nu}_{2}/m^2_{K^{*0}}}{k_2^2-m^2_{K^{*0}}}\frac{1}{q^2-m^2_{D_s^{+}}},\label{eq25}
\end{align}
where the expressions in the curly brackets, $\{1$, $i(k_2^{\alpha}-k_1^{\alpha})\epsilon^{X}_{\alpha}\}$, are for $X_0$ and $X_1$, respectively.
\subsection{Three-body decay width}
Similarily, the decay amplitudes of the triangle diagrams of Fig.~\ref{mkuds}, evaluated in the initial state center of mass frame, are
\begin{align}
{\cal{M}}_{a}(X_J\to\pi^0K^{0}\bar{D}^{*0})&=\frac{ig_hg_{X_J\bar{D}^{*}K^{*}}}{2\sqrt{2}}{\cal{C}}_Y^{I}\Phi[(p_2\omega_{K^{*0}}-q\omega_{\bar{D}^{*0}})^2]\nonumber\\
                                          &\times(p_3-p_1)^{\mu}\{1,i(q-p_2)^{\alpha}\epsilon_{\alpha}(p)\}\nonumber\\
                                          &\times\frac{-g_{\mu\nu}+q_{\mu}q_{\nu}/m^2_{K^{*0}}}{q^2-m^2_{K^{*0}}+im_{K^{*0}}\Gamma_{K^{*0}}}\epsilon^{*\nu}(p_2),\\
{\cal{M}}_{a}(X_J\to\pi^-K^{+}\bar{D}^{*0})&=\frac{ig_hg_{X_J\bar{D}^{*}K^{*}}}{2}{\cal{C}}_Y^{I}\Phi[(p_2\omega_{K^{*0}}-q\omega_{\bar{D}^{*0}})^2]\nonumber\\
                                           &\times(p_3-p_1)^{\mu}\{1,i(q-p_2)^{\alpha}\epsilon_{\alpha}(p)\}\nonumber\\
                                                 &\times\frac{-g_{\mu\nu}+q_{\mu}q_{\nu}/m^2_{K^{*0}}}{q^2-m^2_{K^{*0}}+im_{K^{*0}}\Gamma_{K^{*0}}}\epsilon^{*\nu}(p_2),\\
{\cal{M}}_{b}(X_J\to\pi^0K^{+}\bar{D}^{*-})&=\frac{ig_hg_{X_J\bar{D}^{*}K^{*}}}{2\sqrt{2}}{\cal{C}}_Y^{I}\Phi[(p_2\omega_{K^{*+}}-q\omega_{\bar{D}^{*-}})^2]\nonumber\\
                                          &\times(p_3-p_1)^{\mu}\{1,i(q-p_2)^{\alpha}\epsilon_{\alpha}(p)\}\nonumber\\
                                                 &\times\frac{-g_{\mu\nu}+q_{\mu}q_{\nu}/m^2_{K^{*+}}}{q^2-m^2_{K^{*+}}+im_{K^{*+}}\Gamma_{K^{*+}}}\epsilon^{*\nu}(p_2),\\
{\cal{M}}_{b}(X_J\to\pi^+K^{0}\bar{D}^{*-})&=\frac{ig_hg_{X_J\bar{D}^{*}K^{*}}}{2}{\cal{C}}_Y^{I}\Phi[(p_2\omega_{K^{*+}}-q\omega_{\bar{D}^{*-}})^2]\nonumber\\
                                             &\times(p_3-p_1)^{\mu}\{1,i(q-p_2)^{\alpha}\epsilon_{\alpha}(p)\}\nonumber\\
                                                 &\times\frac{-g_{\mu\nu}+q_{\mu}q_{\nu}/m^2_{K^{*+}}}{q^2-m^2_{K^{*+}}+im_{K^{*+}}\Gamma_{K^{*+}}}\epsilon^{*\nu}(p_2),
\end{align}
where the expressions in the curly brackets, $\{1$, $i(q-p_2)^{\alpha}\epsilon_{\alpha}(p)\}$, are for $X_0$ and $X_1$, respectively.

Once the amplitudes are determined, the corresponding partial decay widths can be easily obtained, which read as,
\begin{align}
d\Gamma(X_J\to\bar{D}K)&=\frac{1}{2J+1}\frac{1}{32\pi^2}\frac{|\vec{p}_1|}{m^2_{X_J}}\bar{|{\cal{M}}|^2}d\Omega,\\
d\Gamma(X_J\to\bar{D}^{*}K\pi)&=\frac{1}{2J+1}\frac{1}{(2\pi)^5}\frac{1}{16m_{X_J}^2}\bar{|{\cal{M}}|^2}|\vec{p}^{*}_3|\nonumber\\
                                  &\times{}|\vec{p}_2|dm_{\pi{}K}d\Omega^{*}_{p_3}d\Omega_{p_2},
\end{align}
where  $J$ is the total angular momentum of the $X_J$,  $|\vec{p}_1|$ is the three-momenta of the decay products in the center
of mass frame, and the overline indicates the sum over the polarization vectors of the final hadrons.  The ($\vec{p}^{*}_3,\Omega^{*}_{p_3}$)
is the momentum and angle of the particle $K$ in the rest frame of $K$ and $\pi$, and $\Omega_{p_2}$ is the angle of the $\bar{D}^{*}$ in
the rest frame of the decaying particle.  The $m_{\pi{}K}$ is the invariant mass for $\pi$ and $K$ and $m_{\pi}+m_{K}\leq{}m_{\pi{}K}\leq{}M-m_{\bar{D}^{*}}$.
The total decay width of the $X_J$ is the sum of $\Gamma(X_J\to{}\bar{D}K)$ and $\Gamma(X_J\to\pi{}K\bar{D}^{*})$.

\section{results and discussions}
\begin{figure}[htbp]
\begin{center}
\includegraphics[scale=0.4]{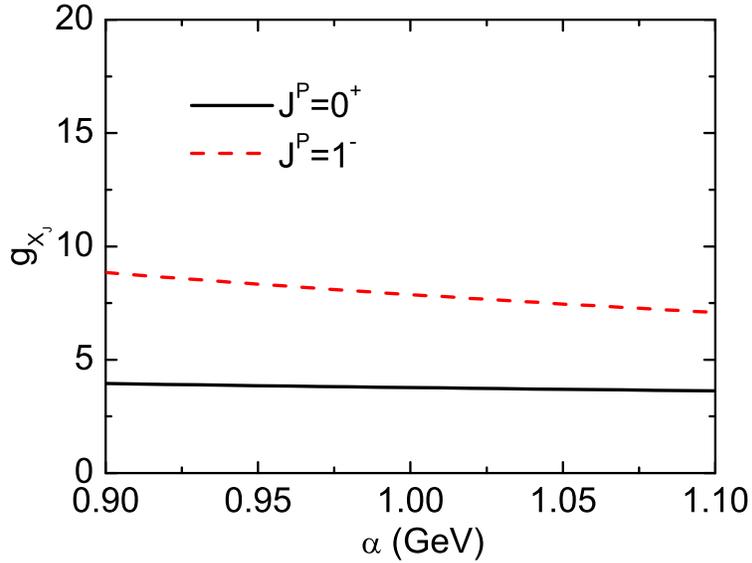}
\caption{Dependences  of the coupling constant of vertex $X_J\bar{D}^{*}K^{*}$  on the parameter $\alpha$ for
different spin-parity assignments.  The coupling constant $g_{X_J}$ for the case of $J^P=0^+$ is in units
of GeV and for the case of $J^P=1^-$ is dimensionless. }\label{mku-coupling}
\end{center}
\end{figure}

 In order to obtain the allowed two body decay widths
through the triangle diagrams shown in Fig.~\ref{mku} and three body decay widths in Fig.~\ref{mkuds}, we first compute the coupling constant
$g_{X_J\bar{D}^{*}K^{*}}$($\equiv{g_{X_J}}$).  With a value of the cutoff $\alpha=0.9-1.1$ GeV, these coupling constants are shown
in Fig~\ref{mku-coupling}.  We note that they decrease slowly with the increase of the cutoff, and the coupling constant
is almost independent of  $\alpha$.  The different $\alpha$ dependences
reflect the different distribution of the two constituents, $\bar{D}^{*}$ and $K^{*}$, in the hadronic molecular $X_J$ states.

\begin{figure}[htbp]
\begin{center}
\includegraphics[scale=0.6]{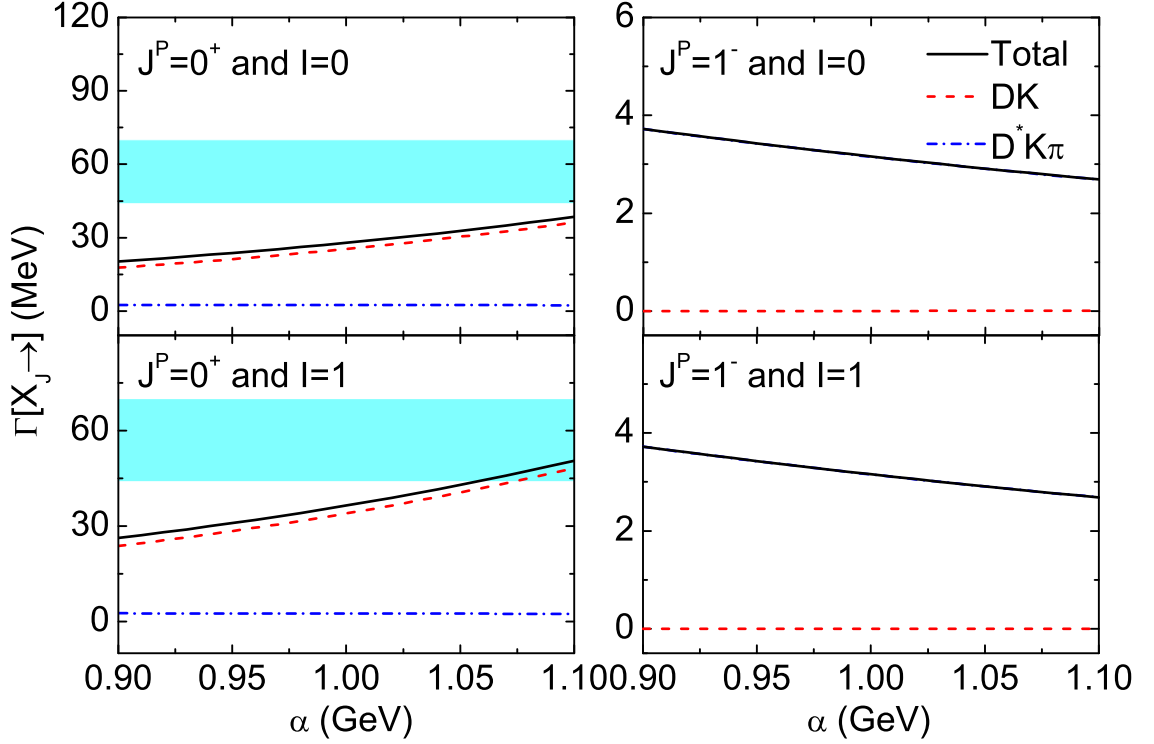}
\caption{Partial decay widths of the $X_J\to{}\bar{D}K$ (dash dashed lines), $X_J\to{}\bar{D}^{*}\pi{}K$ (blue dash dotted lines), and the total decay width (black solid lines) with different spin-parity and isospin assignments for the $X_J$ as a function of the parameter $\alpha$. The oycn error bands correspond to the experimental total decay width~\cite{lhcb-x2900}.}\label{width}
\end{center}
\end{figure}

We show the dependence of the total decay width on the cutoff $\alpha$ in Fig.~\ref{width}.  In the present study,
we vary $\Lambda$ from 0.9 to 1.1 GeV.  In this $\alpha$ range, the total decay width increases for the case of $J^P=0^{+}$,
while it decreases for the $J^P=1^{-}$ case.  The three-body decay widths for both $J^P=0^+,1^-$ and $I=0,1$ are in the range of 2 to 3 MeV, while the two-body decay width for $J^P=0^+$ are at the order of a few tens of MeV, but that for the $J^P=1^-$ are less than 1 MeV (see also Table \ref{tablewidth}). A possible explanation for this is that the width of a $P$-wave molecule is heavily dependent on the spatial distributions of its constituents, as one can see from Eqs.~(\ref{eq22}-\ref{eq25}).

From  Fig.~\ref{width}, we find that the calculated total decay width
for the case of $I(J^P)=1(0^{+})$ is comparable with that of the experimental total width in the range of
$\alpha=1.06-1.1$ GeV, while an even larger $\alpha$ is needed for $I(J^P)=0(0^+)$.  Although a value of $\alpha=1.0$ is preferred based on previous studies ~\cite{Huang:2019qmw,Faessler:2007us,Dong:2009yp,Dong:2009uf,Dong:2017gaw,Xiao:2019mst,Huang:2018wgr,Huang:2018bed}, considering that the fact our results should be considered as the lower limits because it is possible that other decay modes exist, our study did indicate a sizeable $D^*\bar{K}^*$ component in the $X_0$ wave function.
The corresponding partial decay widths of $X_J\to\bar{D}K$, $\bar{D}^{*}\pi{}K$, and the total decay widths for different
spin-parity and isospin assignments of $X_J$ are listed in Tab.~\ref{tablewidth}.  For comparison, we show the
results from the LHCb Collaboration as well~\cite{lhcb-x2900}.   The results show that the $X_0(2866)$ might have a sizeable $D^*\bar{K}^*$ component while the $X_1(2904)$ cannot be explained as a $D^*\bar{K}^*$ molecule. We note that in Ref.~\cite{Karliner:2020vsi}, the $X_0(2866)$ is found to be compatible with a compact tetraquark state.

\begin{table}[htbp!]
\centering
\caption{ Partial decay widths of $X_J\to\bar{D}K$, $\bar{D}^{*}\pi{}K$, and the total decay width for different spin-parity and isospin assignments of $X_J$, in comparison with the LHCb results~\cite{lhcb-x2900}.  Results for the preferred value of $\alpha=1$ GeV are given as central values and the uncertainties originate from the variation of $\alpha$ from 0.9 to 1.1 GeV.  All widths are in units of MeV.}\label{tablewidth}
\begin{tabular}{ccccccccccccccccccccc}
\hline\hline
                          &~~&\multicolumn{2}{c}{$X_0$}       &~~ &\multicolumn{2}{c}{$X_1$}            \\\cline{3-4} \cline{6-7}
 Decay models             &~~& $I=0$    ~~~~~~~~~&$I=1$             &~~ &$I=0$ ~~~~~~~~~& $I=1$                              \\\hline
$\bar{D}K$   &  &$25.42^{-7.71}_{+10.73}$   &$33.95^{-10.25}_{+14.21}$ &&$3.10^{-0.81}_{+0.79}(\times10^{-3})$ &$0.81^{-0.22}_{+0.27}(\times10^{-3})$ \\
$\bar{D}^{*}\pi{}K$       &  &$2.48^{-0.08}_{+0.07}$   &$2.48^{-0.08}_{+0.07}$ &&$3.16^{-0.47}_{+0.56}$&$3.16^{-0.47}_{+0.56}$    \\
Total                &  &$27.90^{-7.79}_{+10.8}$  &$36.43^{-10.33}_{+14.28}$ &     &$3.16^{-0.47}_{+0.56}$&$3.16^{-0.47}_{+0.56}$\\
 Exp.~\cite{lhcb-x2900}                &  &\multicolumn{2}{c}{$57.2\pm{}12.9$} &     &\multicolumn{2}{c}{$110.3\pm{}11.5$}&\\
\hline\hline
\end{tabular}
\end{table}

\section{Summary}
We studied the two-body and three-body strong decays of the two states $X_0(2866)$ and $X_1(2904)$ assuming that
they are bound states of $D^*\bar{K}^*$. The couplings of these states to their components are
fixed by  the Weinberg compositeness condition. The two-body decays are via triangle diagrams with exchanges of a pseudoscalar meson $\pi$, $\eta$, or $D_s$, where the three-body decays happen at tree level. With the other couplings fixed from relevant experimental data, the only remaining parameter is the cutoff $\alpha$. We showed that with the well accepted range of $0.9\sim1.1$ GeV, the so-obtained decay width for the $X_0(2866)$ is in marginal agreement with the LHCb measurement but that for the $X_1(2904)$ is much smaller. As a result, we conclude that the $X_0(2866)$ may have a large $D^*\bar{K}^*$ component (also a non-negligible compact tetraquark component) but the $X_1(2904)$ cannot be of molecular nature.

Such a conclusion is consistent with the OBE model of Ref.~\cite{Liu:2020nil}. We note that a recent study by Karliner and Rosner favors the
explanation of the $X_0$ as a compact tetraquark state~\cite{Karliner:2020vsi}, while the lattice QCD study of Ref.~\cite{Hudspith:2020tdf} found no tetraquark candidate in this channel. As a result, more works are urgently needed to clarify the nature of these latest additions to the family of exotic mesons.

\section*{Acknowledgements}

This work was partly supported the National Natural Science Foundation of China (NSFC) under Grants Nos. 11975041, 11735003, 11961141004, and 11961141012, and the Youth Innovation
Promotion Association CAS (2016367)..

%\end{CJK*}
\end{document}